# UPGRADE OF THE SDL CONTROL SYSTEM

S.K. Feng, W.S. Graves, Y.N. Tang
Brookhaven National Laboratory, Upton, NY 11973, USA

## Abstract

The Source Development Lab (SDL) at BNL consists of a 230 MeV electron linac and a 10 m long wiggler for short wavelength Free Electronic Laser (FEL) development. The original control system [1] was based on the one in use at the National Synchrotron Light Source. In 2000 the control system was upgraded to be based on EPICS to facilitate software interfaces for extensions at all levels. The FEL began commissioning in May 2001. The upgraded control systems have been stable, reliable and easy to operate. An overview of the control systems, including for the linac, the waveform digitizer controls, video system/image processing and radiation monitor/alarms will be presented.

## 1 INTRODUCTION

In SDL, the challenging physics requirements and different favorite approaches for implementations among physicists, users, and programmers initiated the upgrade of the control system to be based on EPICS to facilitate software interfaces for extensions at all levels. As of today, the mixed hardware platforms and software tools of the SDL control systems are integrated homogeneously (see Figure 1).

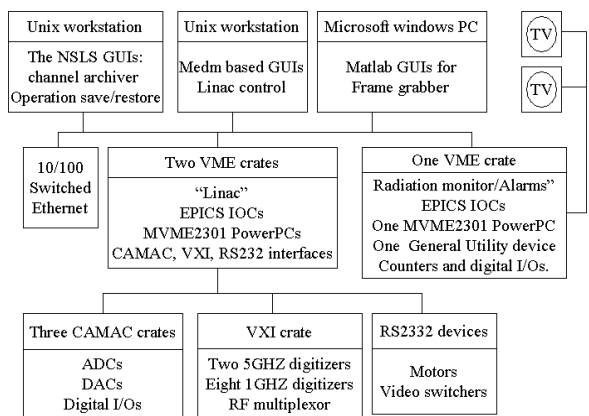

Figure 1: The Hardware/software Illustration of SDL Control System

The VME Input/Output Controllers have interfaces to both CAMAC and VXI crates, and RS232 devices. The software for an RF multiplexor (MUX) and fast VXI waveform digitizers is developed to measure/record/display many voltage waveforms.

## 2 IMPLEMENTATION

### 2.1 The waveform digitizers

In the VXI crates, there are two 4-channel 1 GS/s waveform digitizers used for general purpose data acquisition and one 2-channel 5GS/s waveform digitizer used for beam position monitor (BPM) processing. The VXI waveform digitizer control and display, consisting of a MVME2301 PPC/VME-MXI-II/MXI-VXI-II and VXI scopes hardware and EPICS IOC core in low level control, is developed with the MEDM tool and Cartesian plotting to effectively present the required information on the workstation (see Figure 2).

The requirement of the SDL application is to refresh the display of the ten waveforms simultaneously on the workstation at a rate close to 10HZ for each channel. The challenge is to overcome the limited data transfer capability between the CPU and the waveform digitizer, and to transfer 1K bytes of voltage data in floating point format for each waveform (e.g. 80K bytes of data for the ten channels) across the network to display on the workstation without compromising the performance requirement.

Theoretically, the data move performance of the VME-MXI-II and MXI-VXI-II pair tops 8-10 MB/s at single cycles. However, the benchmark of data move between the MVME2301 PPC and VXI waveform digitizers ranges 16-30 KB/s via the Fast Data Channel (FDC) protocol at single cycles. That limits the readouts of each waveform data (2K bytes) per scope channel to be 8-15 HZ in the low level control.

At an early stage of device driver development with EPICS, the ten channels of scopes performed at 2 Hz of waveform display refresh rate when the records were scanned periodically. After further enhancement of real-time programming by employing EPICS event scan and scope driver interrupts, the ten channels of scopes, while running simultaneously under average network loads, achieved an average of 10 Hz of waveform display refresh rate on two channels and 67HZ of waveform display refresh rate on the other eight channels.

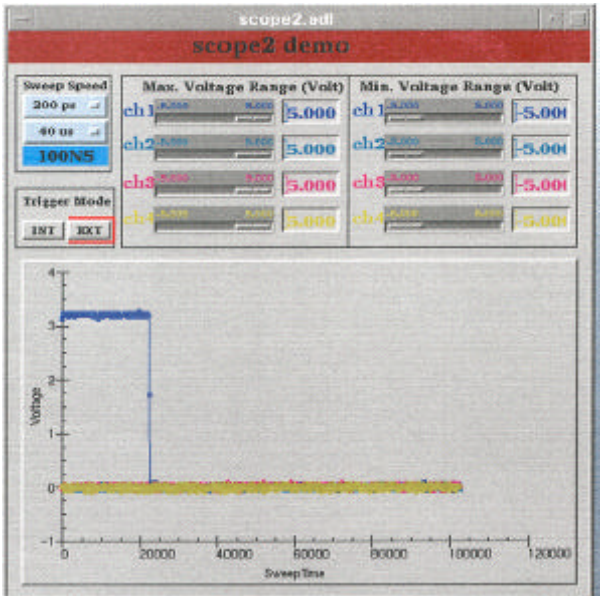

Figure 2: waveform digitizer control/display demo

## 2.2 Linac control systems

The linac control systems consist of two VME IOCs controlled by MVME2301 PowerPCs in low level control. Most of the ADCs, DACs and digital I/O for the linac are carried using legacy CAMAC equipment. Although the VME/CAMAC driver was obtained from the community of the EPICS collaborations, the EPICS device support routines for our CAMAC equipment were developed in house. Stepper motors and video switchers are controlled via RS232. Once device support routines are developed, new records of the supported device can be added to the system by editing the database without the need of changing the low level device drivers and support routines, thanks to the object-oriented structure of the EPICS layers. The EPICS save/restore program embedded in the IOC facilitates the restore of the output parameters in case of an IOC reboot.

The high level linac control is implemented with the MEDM tool of EPICS to provide friendly Graphical User Interfaces (GUIs). Figure 5 shows the main page of the linac GUI control. The motif-based channel archiver (the history program and its graphic display) and save/restore programs for various control operations, which were in use at the NSLS were modified to include the EPICS channel access to communicate with the linac IOCs.

## 2.3 Video System and Image Processing

The primary beam diagnostics for SDL are YAG:Ce scintillators that intercept the electron beam and produce light that is imaged by Cohu 4910 CCD cameras. The RS-170 video signals are fed from approximately 60 cameras through video matrix switchers to monitors and a PC based frame grabber for image processing. The Matrox Meteor II/MC 8 bit analog frame grabber is controlled by Matlab MEX C function calls and includes a GUI interface (Figure 3). The frame grabber and analysis software can set and read EPICS process variables utilizing an ActiveX interface.

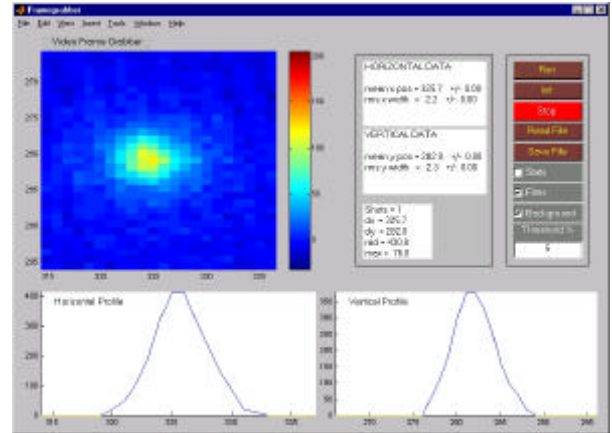

Figure 3: Matlab GUI to PC-based frame grabber

The frame grabber and associated Matlab beam analysis programs provide near real-time diagnostics capability that has been very useful during the commissioning process. Figure 4 shows an emittance measurement that is performed by Matlab function calls. The beam sizes are read from the frame grabber and magnet settings are read from the EPICS database. These values are then used to calculate the emittance and beam Twiss parameters. This enables fast lattice development and optimization of beam parameters.

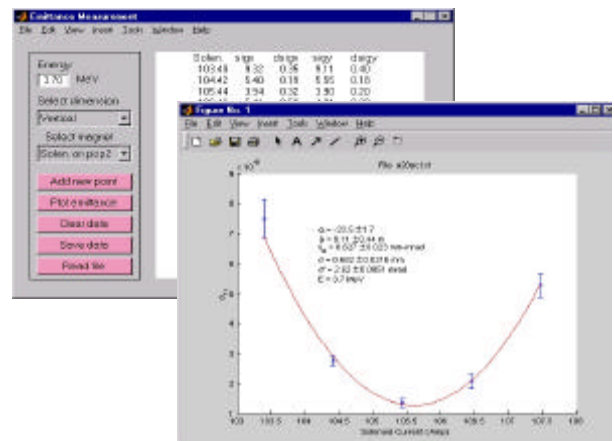

Figure 4: Automated emittance measurement using Matlab and Epics.

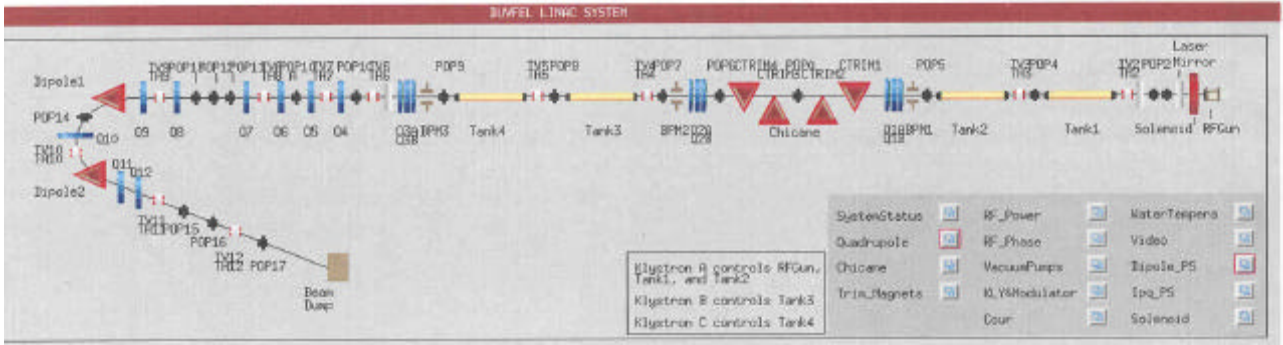

Figure 5: The main page of the SDL control system

## 3 WORK IN PROGRESS

### 3.1 Radiation monitors/alarm status

The radiation monitors/alarm status IOC consists of one MVME2301 PowerPC, three 5 - channel counters, digital I/Os and a general utility board for TV message display [2]. The radiation monitor is required for radiation safety to monitor and display radiation levels and alarm status on the TVs. The system is implemented to the same specification as those of the alarm/radiation system which is in use at the NSLS. The advantages of using the TV display instead of an X window display are, 1) the information is real-time independent of the network status, and 2) it eliminates the need of point and click and shuffling the screen. Thus, there is no need to train users who are not familiar with X-windows environment.

We decided it is worthwhile to rewrite the same specifications to be based on EPICS to facilitate software interfaces for extensions at all levels.

## 4 FUTURE PLANS

For the further development of the SDL control system, we would like to use the VME64x crate, IP carriers and IP modules.

For the image processing, we would like to use the same PC-based frame grabber to be controlled under the LINUX operating system as one of the EPICS IOCs, which facilitates high level GUI remote controls via a workstation.

The workstation for VxWorks development and high level GUI control will be replaced by a high-end PC running the LINUX operating system due to concern for cost and performance.

## 5 CONCLUSIONS

The SDL control systems have been stable, reliable and easy to operate. We have been able to utilize the pre-existing hardware to minimize costs. We benefited from the object-oriented structure of the EPICS layers to reduce the efforts of maintenance and expansions without compromising the performance of low level real-time control (e.g. the implementations of the waveform digitizers ). The upgrade is satisfactory that we will continue to develop and expand our existing control system to increase functionality, ease of use and reliability of the facility.

## 6 ACKNOWLEDGEMENTS

The authors wish to extend their thanks to J.D. Smith and Eric Blum of BNL and all the other people who have helped with the integration of the SDL control system. This work performed under DOE contract DE-AC02-76CH00016.

## REFERENCES

[1] W.S. Graves, S.K. Feng, P.S. Pearson, J.D. Smith, "Innovative Aspects of the SDL control System", Proc. PAC97 (Vancouver, 12-16 May 1997).
[2] S. Ramamoorthy, J.D.Smith, "GPLS VME Module: A Diagnostic and Display Tool for NSLS Micro Systems", Proc. PAC99 (New York City, 29 March – 2 April 1999).